# Topological phase transitions driven by magnetic phase transitions in $Fe_xBi_2Te_3$ ($0 \leq x \leq 0.1$) single crystals

Heon-Jung Kim[1,*], Ki-Seok Kim[2], J.-F. Wang[3], V. A. Kulbachinskii[4], K. Ogawa[5], M. Sasaki[6], A. Ohnishi[6], M. Kitaura[6], Y.-Y. Wu[3], L. Li[3], I. Yamamoto[5], J. Azuma[5], and M. Kamada[5], V. Dobrosavljević[7]

[1] Department of Physics, College of Natural Science, Daegu University, Gyeongbuk 712-714, Republic of Korea

[2] Department of Physics, Pohang University of Science and Technology, Pohang, Gyeongbuk 790-784, Republic of Korea

[3] Wuhan National High Magnetic Field Center, Huazhong University of Science and Technology, Wuhan 430074, China

[4] Physics Department, Moscow State University, Moscow 119899, Russia

[5] Synchrotron Light Application Center, Saga University, Honjo, Saga 840-8502, Japan

[6] Department of Physics, Faculty of Science, Yamagata University, Kojirakawa, Yamagata 990-8560

[7] Department of Physics and National High Magnetic Field Laboratory, Florida State University, Tallahassee, Florida 32306, USA

E-mail: hjkim76@daegu.ac.kr

**Abstract**

We propose a phase diagram for $Fe_xBi_2Te_3$ ($0 \leq x \leq 0.1$) single crystals, which belong to a class of magnetically bulk-doped topological insulators. The evolution of magnetic correlations from "ferromagnetic"- to "antiferromagnetic"- gives rise to topological phase transitions, where the paramagnetic topological insulator of $Bi_2Te_3$ turns into a band insulator with ferromagnetic-cluster glassy behaviours around $x \sim 0.025$, and it further evolves to a topological insulator with valence-bond glassy behaviours, which spans over the region between $x \sim 0.03$ up to $x \sim 0.1$. This phase diagram is verified by measuring magnetization, magnetotransport, and angle-resolved photoemission spectra with theoretical discussions.

---

[*] Corresponding author

The characteristic features of topological insulators originate from the existence of topologically protected gapless surface states [1-4], which gives rise to anomalous transport phenomena [5]. In particular, an anomalous Hall effect of the topological origin and extremely large magnetoresistance (MR) have been observed, attributed to surface Dirac electrons [6]. For the scientific understanding and practical applications, such anomalous transport needs to be controlled, for example by creating magnetic correlations in topological insulators. In fact, several surface- and bulk-doping studies [7-11] reported that surface-doped magnetic impurities are aligned ferromagnetically to cause an excitation gap for surface Dirac electrons [7,8]. However, the gap opening was not always observed [9-11], which suggests that Ruderman-Kittel-Kasuya-Yosida (RKKY) interactions by surface Dirac electrons may be more complicated than expected. Indeed, such RKKY interactions consist of Heisenberg-like, Ising-like, and Dzyaloshinskii-Moriya (DM)-like terms, and are expected to cause frustration for spin dynamics [12-14]. As a result, various spin orders are expected to occur due to interplay between the randomness and spin-orbit interaction provided the chemical potential lies away from the Dirac point of the surface band. This aspect motivated us to dope magnetic ions into the bulk of a topological insulator, controlling anomalous transport phenomena.

In this letter, we examine the effects of various spin orders on topological properties in $Fe_xBi_2Te_3$ ($0 \leq x \leq 0.1$) single crystals by measuring the MR, Hall resistance, magnetization, and angle-resolved photoemission spectroscopy (ARPES). Transport properties change drastically, following the evolution of magnetic correlations in the bulk. The Curie-Weiss temperature $\theta$ initially increases with $x$, reaching a maximum at $x = 0.025$. This indicates enhancement of ferromagnetic correlations around $x = 0.025$. Such predominant ferromagnetic correlations turn out to reflect more complicated spin dynamics associated with random positions of magnetic ions. Ferromagnetic-cluster glassy behaviours are observed around this $x$ value along with drastic changes in the MR and Hall effect. The characteristic features of topological-insulator samples, such as large MR and anomalous Hall effect are no longer observed in the $Fe_{0.025}Bi_2Te_3$ sample, which suggests that a gap opens at the Dirac point of the surface band. Indeed, we observe the gap opening from our ARPES measurement. Interestingly, $Fe_xBi_2Te_3$ samples with $x = 0.05$ and $0.1$ exhibit essentially the same transport behaviour as undoped $Bi_2Te_3$ without any sign of "ferromagnetism". Furthermore, ARPES fails to detect any gap opening at the Dirac point in the $Fe_{0.1}Bi_2Te_3$ sample, confirming the recovery of the topologically nontrivial nature. This puzzling observation of the re-entrant behaviour is resolved, considering that the Weiss temperature $\theta$ becomes negative above $x = 0.025$, where antiferromagnetic correlations become more dominant than ferromagnetic interactions, giving rise to valence-bond glassy behaviours. The entire result is summarized in the phase diagram of Fig. 1.

The powder x-ray diffraction (XRD) data, measured by using crushed $Fe_xBi_2Te_3$ crystals with $x=0.0, 0.025, 0.05, 0.075$, and $0.1$ are shown in Fig. 2(a). The XRD neither showed any significant change of the $Bi_2Te_3$ structure nor exhibited the appearance of superstructures. Figure 2(b) shows the $M(T)$ curves of the $Fe_xBi_2Te_3$ single crystals under magnetic fields perpendicular to the cleaved (111) plane. In addition, the inset in Fig. 2(b) displays $1/M(T)$ for the samples with x=0.0125, 0.05, and 0.1. Clearly, these curves are linear in the high temperature region, which indicates that they follow the Curie-Weiss law. At lower temperatures, the $M(T)$ curves deviate from this linearity, the temperature of which depends on $x$ as spin correlations develop. In particular, the sample with $x = 0.025$ shows a very clear deviation or a hump at approximately $T = 120$ K in the original $M(T)$ curve. To clarify the nature of this behaviour, the $M(T)$ curves were fitted to the Curie-Weiss formula, $M(T)/H = C/(T + \theta)$ with a constant $C = N\mu_{eff}^2/(3k_B)$, where $N$ is the number of impurities, $\mu_{eff}$ is the effective magnetic moment, and $k_B$ is the Boltzmann constant. The inset of Fig. 2(b) shows our results of fitting, plotted as (red) dotted lines. The Curie-Weiss temperatures θ of the samples with x=0.0125 and 0.05 are positive and negative, respectively, while it becomes almost zero in the sample with x=0.1. This suggests that average magnetic interactions change around x=0.025 from ferromagnetic to antiferromagnetic correlations.

To understand the change of magnetic correlations in more detail, we present the Weiss temperature and effective magnetic moment in Fig. 2(c), determined from the Curie-Weiss fitting to the $M(T)$ curves as a function of $x$. For $x \leq 0.025$, the Weiss temperature increases with x, reaching a maximum at $x = 0.025$. The maximum Weiss temperature suggests that ferromagnetic correlations are predominant around $x = 0.025$. On the other hand, such average magnetic correlations change drastically from ferromagnetic- to antiferromagnetic- interactions around $x = 0.03$. The negative θ for $x > 0.025$ is attributed to the enhancement of antiferromagnetic correlations. The average magnetic moment also shows a sudden decrease just above $x\sim0.025$, again indicating that predominant magnetic correlations change from ferromagnetic- to antiferromangetic- across this particular x value.

We would like to emphasize that such average magnetic correlations, reflected in Curie-Weiss temperatures and magnetic moments, hide more complicated spin dynamics. Although the susceptibility curve for x = 0.0125 increases monotonically, the fact that it saturates to a finite value at $T = 0$ implies that not only ferromagnetic correlations but also antiferromagnetic interactions play their certain roles in this random system of magnetic impurities, giving rise to glassy behaviours. The susceptibility curve for x = 0.025 shows more complicated spin dynamics. The cusp around 130 K seems to indicate that random magnetic interactions are at work. In addition, the continuous increase with a power-law behaviour but the change of its slope around 15 K and the saturation of such divergent behaviours at the lowest temperature imply that magnetic correlations evolve from quantum Griffiths behaviours to ferromagnetic-cluster glassy structures. If a magnetically doped system shows about 70% ferromagnetic interactions and 30% antiferromagnetic correlations due to randomly distributed positions of magnetic ions, it is natural to expect that ferromagnetic clusters are formed and their random spin correlations can result in glassy behaviours through intermediate complex spin dynamics.

Actually, quantum Griffiths behaviours can be verified by the measurement of the exponent in the temperature dependence of magnetization, $M \sim T^{-\alpha}$. A broad distribution of sizes of such clusters and their effective interactions to other clusters has been claimed to cause the exponent α to be less than 1 [15]. Indeed, we observe such spin dynamics as shown in Fig. 2(d). Just below the cusp, the value of α is around 1. On the other hand, it becomes much reduced to 0.163 from approximately 15 K, indicating the signature of the quantum Griffiths phase [16]. As temperature decreases further, these ferromagnetic clusters become frozen. As a result, the divergent behaviour weakens, the spin susceptibility being saturated to a finite value. As the concentration of magnetic impurities increases, predominant magnetic correlations evolve from ferromagnetic- to antiferromagnetic-, verified from the Curie-Weiss plot at x = 0.05 and 0.1, discussed before. Considering that the spin susceptibility increases but saturates at low temperatures, we speculate that the ground state is a sort of "spin liquid", which may correspond to valence-bond glassy behaviours.

The observed change in the spin dynamics profoundly affects the magnetotransport properties of $Fe_xBi_2Te_3$ single crystals. Figure 3 shows the MR (= $\Delta\rho/\rho_0$) and Hall resistance of $Bi_2Te_3$, $Fe_{0.025}Bi_2Te_3$, and $Fe_{0.1}Bi_2Te_3$ single crystals at 4.2 K. The MR and Hall resistance of $Bi_2Te_3$ are typical of the as-grown $Bi_2Te_3$ single crystals as reported previously [6]. The observed MR is ~ 100 % at 4 T and the Hall resistance is nonlinear due to anomalous Hall effects from Berry phase, side jump, and skew scattering contributions [6]. Because the $Bi_2Te_3$ in this study is an as-grown sample, the bulk conduction channel still exists with *p*-type charge carrier. As shown in Fig. 3, there is no qualitative difference between $Bi_2Te_3$ and $Fe_{0.1}Bi_2Te_3$. It should be noted that both the MR and Hall resistance of the samples with x=0, 0.0125, 0.5, 0.075, and 0.1 are scaled by the proper $MR^*$'s and $H^*$'s as shown in Fig. 3(c) and (d), respectively, in which $H^*$ is a characteristic field characterizing surface conduction due to Dirac fermions and $MR^*$ is the MR at $H = H^*$. Therefore, the features related to the surface conduction, such as large MR and anomalous Hall effect, are still observable in $Fe_{0.1}Bi_2Te_3$, suggesting the survival of surface conduction.

In contrast, the MR and Hall resistance of $Fe_{0.025}Bi_2Te_3$ are quite different and completely conventional in that the MR and Hall resistance are quadratic and linear with *H* up to 4 T, respectively. The magnitude of MR, which is approximately 8 % at 4 T, is reduced drastically compared to $Bi_2Te_3$ and $Fe_{0.1}Bi_2Te_3$. This sample simply follows the Boltzmann transport theory. Therefore, the surface conduction by Dirac fermions appears to be suppressed completely. In addition, the hole mobility $\mu$ and the hole number $p$ estimated from the quadratic MR and linear Hall resistance are 0.14 m$^2$/Vs and $1.8 \times 10^{19}$ cm$^{-3}$, respectively, which are in a range of conventional doped semiconductors. This also supports the bulk electrical conduction in this sample. Predominant ferromagnetic correlations and conventional behaviours of electrical transport properties observed at the $x = 0.025$ samples are quite correlated.

The changes of the magnetic and transport properties are accompanied by the change of the surface electronic states. Figure 4(a)-(f) are the photoemission intensities and their second derivatives of Bi$_2$Te$_3$, Fe$_{0.025}$Bi$_2$Te$_3$, and Fe$_{0.1}$Bi$_2$Te$_3$, respectively along the Γ-M lines. The Fermi levels of these ARPES spectra were observed to change with time, which seems to be a feature of the Bi$_2$Te$_3$ surface not related to the intrinsic band structure [17]. Because of this, the Fermi levels of the spectra are different. To resolve the dispersions of the surface Dirac bands clearly, the peak positions are determined from the momentum distribution curves (MDCs) by fitting. These peak positions at different energies give the dashed lines in Fig. 4(a)-(f). In Bi$_2$Te$_3$ and Fe$_{0.1}$Bi$_2$Te$_3$, the dispersions are linear near the Dirac point virtually with no gap, implying the existence of the surface Dirac states. On the other hand, in Fe$_{0.025}$Bi$_2$Te$_3$, the dispersion has a gap of 30 ~ 40 meV, suggesting a time-reversal-symmetry broken surface state. The gap opening at the Dirac point in Fe$_{0.025}$Bi$_2$Te$_3$ was also observed in our laser-ARPES experiments. Figure 4(g) and (h) present the stacks of energy distribution curves (EDCs) of Bi$_2$Te$_3$ and Fe$_{0.025}$Bi$_2$Te$_3$, respectively. The laser-ARPES also demonstrates the linear dispersion near the Dirac point in Bi$_2$Te$_3$. In contrast, the surface bands in Fe$_{0.025}$Bi$_2$Te$_3$ have a gap of 30 ~ 40 meV, consistent with the synchrotron-based ARPES results.

Our experiments reported one type of a magnetic phase transition from a ferromagnetic-cluster glassy state to a valence-bond glassy phase, where predominant magnetic correlations change from ferromagnetic- to antiferromagnetic- and at least two topological phase transitions. The first occurs in the region where predominant ferromagnetic correlations exist while the second seems to appear at the magnetic phase transition.

The magnetic phase transition itself is not much unexpected because RKKY interactions between doped magnetic impurities can change from ferromagnetic to antiferromagnetic, depending on their distances. Although a band-structure calculation is needed to understand the nature of the RKKY interactions more accurately, one can estimate the order of magnitude for the critical concentration of magnetic ions that corresponds to a change in the sign of the RKKY interaction. Because the RKKY interaction oscillates on the length scale of $1/2k_F$, a sign change occurs when the number of the magnetic impurities becomes comparable with $L^3 \cdot (2k_F)^3$, where $L$ is the lateral size of a sample. Taking the critical concentration of $x_c \sim 0.025$ with simple algebra produces $k_F \approx 10^7 - 10^8$

cm$^{-1}$, which corresponds to the Fermi energy $E_F \sim 0.01 \sim 0.1$ eV with the effective mass of Bi$_2$Te$_3$. This is in agreement with conventional values estimated by ARPES and de Haas-van Alphen experiments [18].

Two kinds of topological phase transitions can be understood in the following way. The first topological phase transition in the region of $x < 0.025$ is driven by ferromagnetic-cluster glassy behaviours, where magnetic phase transitions are not accompanied. We suggest an effective free energy as a phenomenological model for this topological phase transition,

$$F_{eff} \approx -T \int d\vec{\Phi} P[\vec{\Phi};T] \ln \int D\psi \exp\left[-\int_0^{1/T} d\tau \int \frac{d^3\vec{k}}{(2\pi)^3} \psi_{\vec{k}}^+ \left\{(\partial_\tau - \mu)I \otimes I + H_{\vec{k}}\right\} \psi_{\vec{k}} - \int_0^{1/T} d\tau \int d^3\vec{r} J_{eff} \psi_{\vec{r}}^+ (\vec{\sigma} \otimes I) \psi_{\vec{r}} \cdot \vec{\Phi}\right]$$

with an effective 4 × 4 Dirac Hamiltonian for topological insulators, $H_{\vec{k}} = V_D \vec{k} \cdot \vec{\sigma} \otimes \tau^z + M(k) I \otimes \tau^x$, where $\vec{\sigma}$ and $\vec{\tau}$ represent the spin and orbital indices, respectively. $V_D$ is the velocity of the bulk Dirac electrons and $M(k) = m - \rho k^2$ is an effective mass parameter to incorporate a correct scheme for regularization, both of which can be derived from $k \cdot p$ theory, well-known in the semiconductor physics community. Then, it is straightforward to see the criterion for the topological phase transition in the absence of magnetic impurities, where $m\rho > 0$ ( $m\rho < 0$ ) corresponds to a topological (band) insulator. $\vec{\Phi}$ represents an effective magnetic field originating from ferromagnetic clusters. These ferromagnetic clusters couple to the spin of bulk Dirac electrons with an effective coupling constant $J_{eff}$, whose strength will depend on the size of a ferromagnetic cluster and the distance between clusters.

We introduce a distribution function for the magnetic field of ferromagnetic clusters, which satisfies two conditions. Since ferromagnetic ordering was not observed in our experiments, we require $\langle \vec{\Phi} \rangle = \int d\vec{\Phi} P[\vec{\Phi};T] \vec{\Phi} = 0$. In addition, the distribution function should reproduce the spin susceptibility of our experiment, given by $\chi(T) = \langle \vec{\Phi} \cdot \vec{\Phi} \rangle = \int d\vec{\Phi} P[\vec{\Phi};T] \vec{\Phi} \cdot \vec{\Phi}$. For example, if we assume the Gaussian distribution for the effective magnetic field, it is possible to determine its variance with a zero mean value as a function of temperature. One may regard $P[\vec{\Phi};T]$ as a transformed probability from $\chi(T) = \int dT_{CW} \frac{P[T_{CW}]}{T + T_{CW}}$, where $T_{CW}$ is the Curie-Weiss temperature and $P[T_{CW}]$ is its distribution function. It is straightforward to obtain

$$F_{eff} \approx -T \int d\vec{\Phi} P[\vec{\Phi};T] \int \frac{d^3\vec{k}}{(2\pi)^3} \sum_{s,p=\pm} \ln\left\{1+\exp\left(-s\frac{E_p(\vec{k};\vec{\Phi})}{T}\right)\right\}$$ 
with the energy spectrum of 

$$E_p(\vec{k};\vec{\Phi}) = \sqrt{V_D^2(k_x^2+k_y^2)+\left(J_{eff}\Phi+p\sqrt{M^2(k)+V_D^2 k_z^2}\right)^2}$$, where an Ising anisotropy ($\vec{\Phi}=\Phi\hat{z}$) is assumed for simplicity. It has been shown that a topological phase transition occurs from a topological insulator to a Weyl semi-metal phase, increasing the effective field which corresponds to $J_{eff}\Phi$ [19]. Since the effective field generated from ferromagnetic clusters has its broad distribution, which can be determined from the susceptibility data, we expect that topological insulating states, Weyl semi-metal phases, and band insulating states will form inhomogeneous islands, depending on the strength of the effective magnetic field. A percolation picture among such islands may serve the underlying mechanism of the topological phase transition. However, we note that the classical percolation transition may not coincide with the topological phase transition because of "edge" states of the topological insulating island and quantum interference effects in the presence of spin-orbit interactions.

The second topological phase transition appears to be driven by the magnetic phase transition from a cluster glassy state to a valence bond glassy phase, where predominant ferromagnetic interactions evolve into antiferromagnetic correlations. In this respect it is not so surprising that the topological insulating state is essentially recovered. Actually, the coexistence between antiferromagnetism and topological insulating properties, which allows gapless surface Dirac fermions has been discussed in previous studies [20]. Then, it is natural that the topological insulating state is recovered in the presence of valence bond singlets, where time reversal symmetry is respected "more than" the case of an antiferromagnetic order.

However, various aspects of this phase transition still remain unresolved. In particular, the nature of the magnetic phase transition is not clarified yet. Recalling that the presence of spin-orbit interactions gives rise to frustrating interactions, exotic glassy ordering such as chiral glassy behaviours may occur between ferromagnetic-cluster glassy and valence-bond glassy phases, where the presence of DM interactions favours non-coplanar spin ordering [21, 22]. The magnetic phase transition may also be of the first order, resulting from the abrupt change of magnetic configurations. Then, following this magnetic phase transition, the second topological phase transition can be of the first order. The inhomogeneity in this topological phase transition may not be as important as the first one.

The present study may shed light on the previous controversial results about gap opening of the surface Dirac band. Recall that Ref. [8] reported the gap opening, while the later ARPES studies of Refs. [9,11] claimed the opposite with no virtual difference between magnetic and nonmagnetic ions. These null results are more consistent with the reported positions of the Fermi level far from the

Dirac point, which favours ferromagnetism with in-plane moments, not causing the gap to open. In this respect, the bulk-doping of magnetic ions is more effective than the surface-doping for controlling the topological characters. Indeed, the gapped surface state was realized by magnetically bulk-doping [7]. Our suggested phase diagram is quite general and it can also explain other cases such as Fe doped $Bi_2Se_3$ [7] and $Bi_{2-x}Mn_xTe_3$ [23], where the ferromagnetic "insulating" region is more expanded than the present case. The phase boundary and the area of each phase in Fig. 1 will be determined by the periodicity of the RKKY interaction, given by $1/2k_F$. We suspect that our doped samples are more metallic than those of the previous works. Therefore, in our case the Fermi momentum $k_F$ is relatively larger and the ferromagnetic-interaction dominated region is smaller, allowing us to observe topological phase transitions.

In conclusion, our experiments verified magnetically controlled topological phase transitions by doping magnetic ions into topological insulators. The transport properties of both MR and Hall become normal at low concentrations ($x \leq x_c$) around ferromagnetic-cluster glassy behaviours. At high concentrations ($x \geq x_c$), they turn abnormal, essentially identical to those of topological insulators, when antiferromagnetic correlations are predominant. A phase diagram of $Fe_xBi_2Te_3$ ($0 \leq x \leq 0.1$) single crystals was proposed, based on the magnetization, transport measurements, ARPES, and theoretical discussions. The present study casts a new theoretical challenge, in particular, on how to characterize or define topological phase transitions in the presence of randomly distributed magnetic clusters. This conceptual framework generalizes the physics of dilute magnetic semiconductors [30], introducing topological aspects of electronic structures.

Acknowledgements
This research is supported by Basic Science Research Program through the National Research Foundation of Korea (NRF) funded by the Ministry of Education, Science, and Technology (No. 2012-0007294). KS was supported by the National Research Foundation of Korea (NRF) grant funded by the Korea government (MEST) (No. 2012000550). VD was supported by the NSF Grant No. DMR-1005751. MS expresses his thanks to Prof. T. Iwata and K. Tomita for valuable discussions and their support.

Figure captions

Fig. 1 Phase diagram and topological phase transitions of $Fe_xBi_2Te_3$.

Fig. 2 (a) Powder X-ray diffraction data of $Fe_xBi_2Te_3$ samples with $x$ = 0.0, 0.025, 0.05, 0.075, and 0.1. The upper and lower insets show the $c$ and $a$ parameters, respectively as a function of Fe concentrations. The $c$ parameter slightly increases with increasing $x$ while the $a$ parameter are almost unchanged. (b) Temperature dependence of *M/H* for $Fe_xBi_2Te_3$ with x=0.0125, 0.025, 0.5, and 0.1 for the magnetic fields perpendicular (111) plane. The inset shows 1/*M* curves with Curie-Weiss fitting. (c) The Weiss temperature θ or Curie temperature $T_c$ and the magnetic moment μ determined from the Curie-Weiss fitting as a function of x. The open (closed) circles are the Weiss temperatures for the magnetic field parallel (perpendicular) to the (111) plane. The open (closed) squares are the magnetic moments determined for the magnetic field parallel (perpendicular) to the (111) plane. (d) The log-log plot of the *M(T)* curve for x = 0.025, along with the linear fits at high and low temperatures.

Fig. 3 (a) The magnetoresistance (MR) and (b) Hall resistance as a function of *H* for *p*-doped $Bi_2Te_3$, $Fe_{0.025}Bi_2Te_3$, and $Fe_{0.1}Bi_2Te_3$ single crystals. The scaling behaviors of (c) MR and (d) Hall resistance for $Fe_xBi_2Te_3$ with $x$ = 0, 0.0125, 0.05, 0.0725, and 0.1.

Fig. 4 The photoemission intensity plots of (a) $Bi_2Te_3$, (c) $Fe_{0.025}Bi_2Te_3$, and (e) $Fe_{0.1}Bi_2Te_3$, taken by using a synchrotron along the Γ-M lines. The dashed lines are the fitting curves, determined from the momentum distribution curves. (b), (d), and (f) are their second derivatives. The energy distribution curves of (g) $Bi_2Te_3$ and (h) $Fe_{0.025}Bi_2Te_3$ along the Γ-M lines, measured using the VUV laser source.

**Fig. 1**

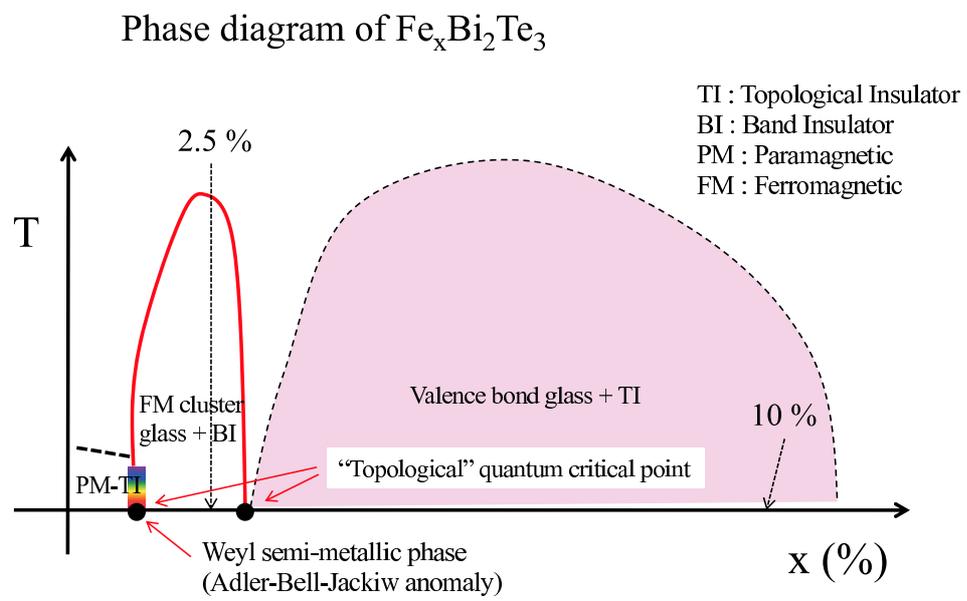

Figure 1

**Fig. 2**

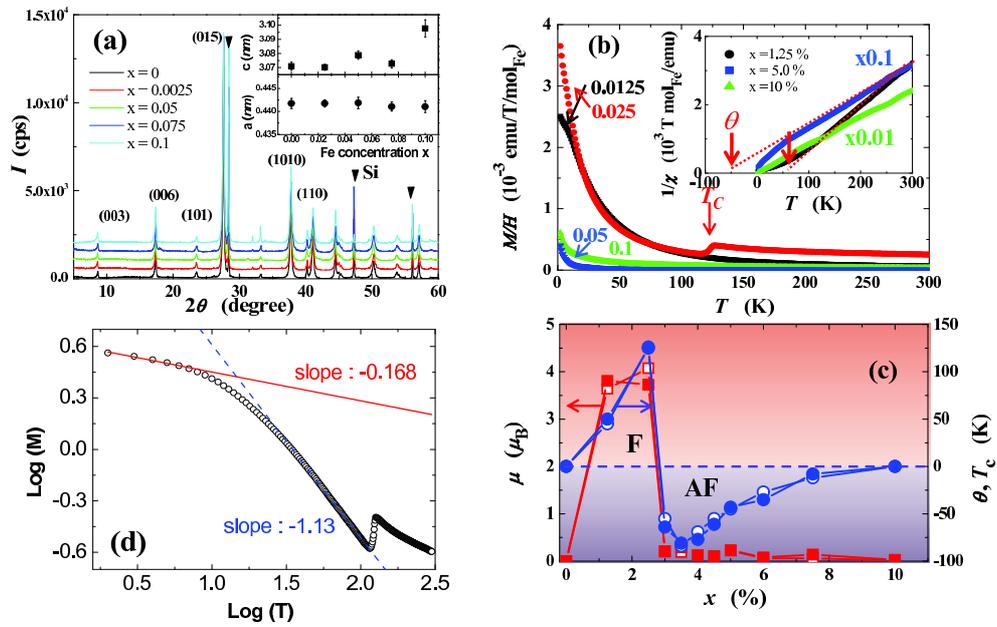

Figure 2

**Fig. 3**

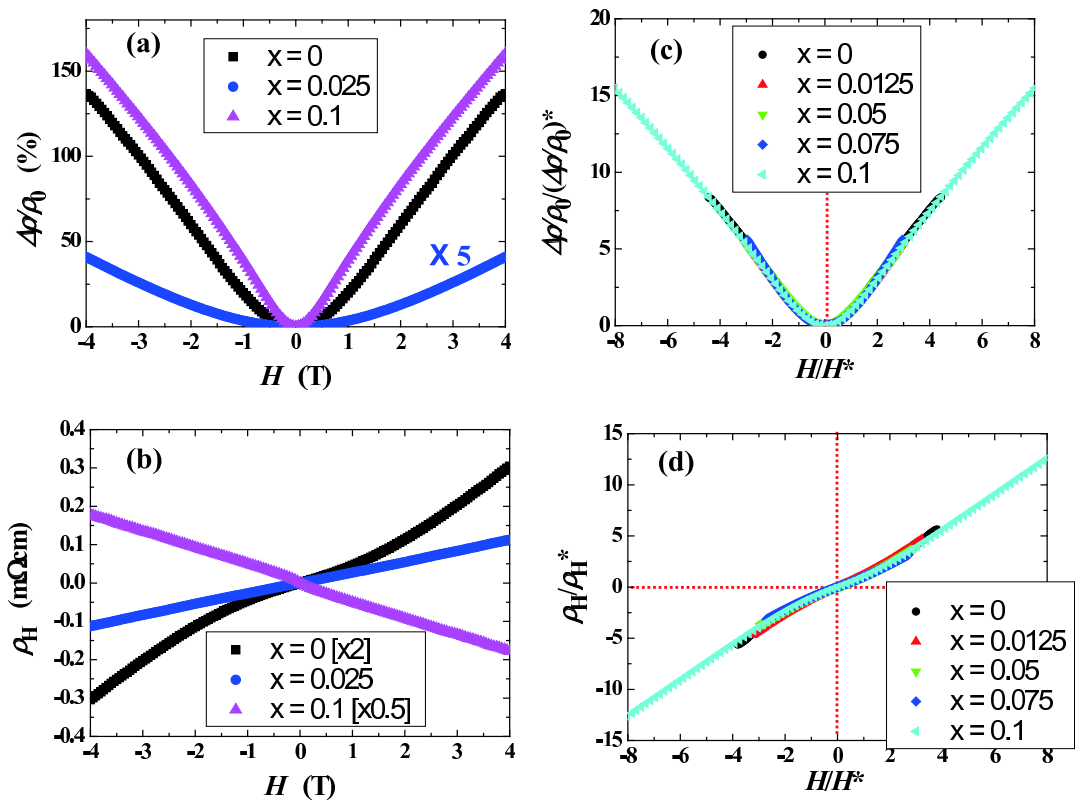

Figure 3

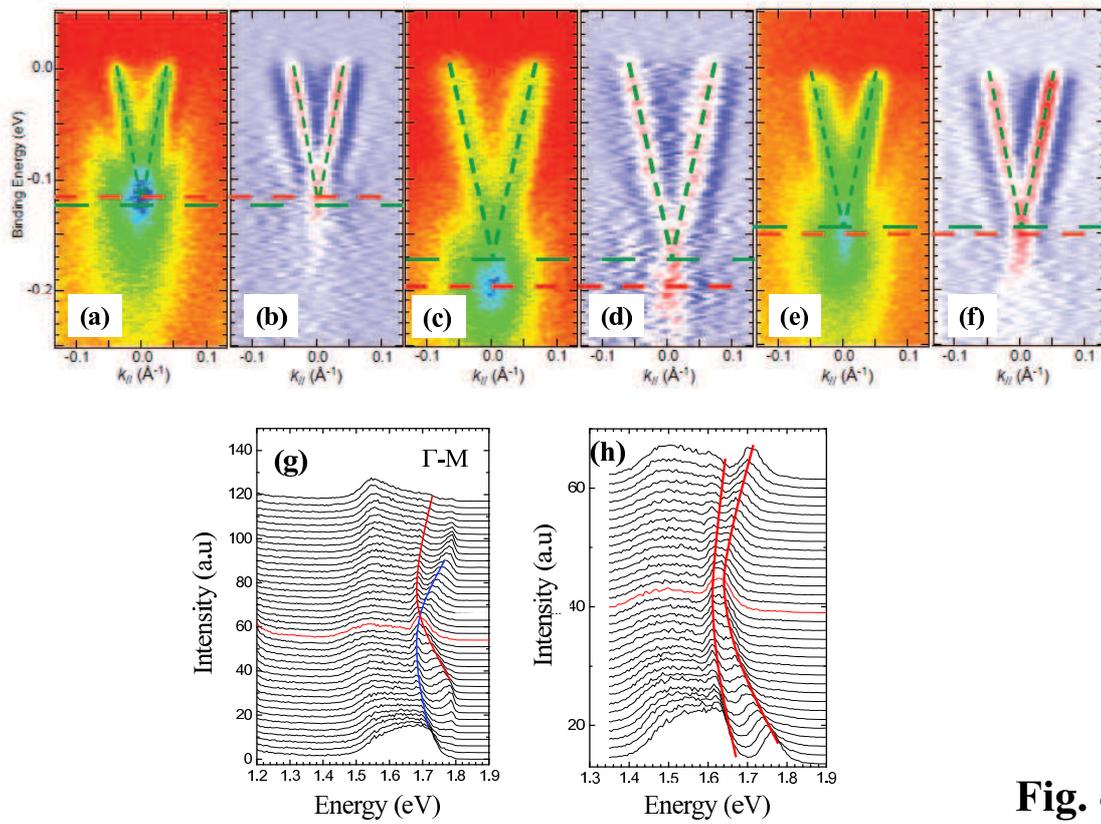

Figure 4